\documentclass{article}

\usepackage{arxiv}

\usepackage[utf8]{inputenc} 
\usepackage[T1]{fontenc}    
\usepackage{url}            
\usepackage{booktabs}       
\usepackage{amsfonts}       
\usepackage{nicefrac}       
\usepackage{microtype}      

\usepackage{mathtools}

\usepackage{graphicx}

\usepackage{caption}

\usepackage{subcaption}

\usepackage{amssymb}
\usepackage{xcolor}

\usepackage{algorithmic}

\usepackage{algorithm} 

\usepackage{hyperref}
\usepackage{doi}
\usepackage{eso-pic}
\newtheorem{prop}{Proposition}

\title{Cryptanalysis of the SIMON Cypher Using Neo4j}

\author{Jonathan Cook*,
        Sabih ur Rehman*
        and M. Arif Khan*\\ 
        {*School of Computing, Mathematics and Engineering, Charles Sturt University, Australia}}

\begin{document}
\maketitle

\begin{abstract}
	The exponential growth in the number of Internet of Things (IoT) devices has seen the introduction of several Lightweight Encryption Algorithms (LEA). While LEAs are designed to enhance the integrity, privacy and security of data collected and transmitted by IoT devices, it is hazardous to assume that all LEAs are secure and exhibit similar levels of protection. To improve encryption strength, cryptanalysts and algorithm designers routinely probe LEAs using various cryptanalysis techniques to identify vulnerabilities and limitations of LEAs. Despite recent improvements in the efficiency of cryptanalysis utilising heuristic methods and a Partial Difference Distribution Table (PDDT), the process remains inefficient, with the random nature of the heuristic inhibiting reproducible results. However, the use of a PDDT presents opportunities to identify relationships between differentials utilising knowledge graphs, leading to the identification of efficient paths throughout the PDDT. This paper introduces the novel use of knowledge graphs to identify intricate relationships between differentials in the SIMON LEA, allowing for the identification of optimal paths throughout the differentials, and increasing the effectiveness of the differential security analyses of SIMON. 
\end{abstract}


\section{Introduction}  \label{Sec:Introduction}
Significant security challenges confront Internet of Things (IoT) devices due to the lack of built-in encryption \cite{meneghello2019iot}. An absence of encryption on the device can result in substantial security risks to the data collected and transmitted by such devices leading to the loss of assets, credentials, and confidentiality \cite{ali2018cyber}. Incorporating encryption on physical IoT devices will enhance the integrity of data collected and transmitted by IoT devices. However, most IoT devices have processing power constraints that limit their ability to run complex encryption algorithms \cite{cook2023security}. To address this limitation, the United States National Security Agency (NSA) developed two LEAs in 2013 suitable for resource-constrained devices \cite{beaulieu2013simon}. Although these two encryption algorithms are not supported by the National Institute of Standards and Technology or the International Organisation for Standardisation \cite{ashur2021account}, they continue to elicit considerable interest from the research community. Researchers are particularly interested in the security analysis of encryption algorithms, particularly SIMON.

A security analysis of encryption algorithms is an essential stage in their development \cite{de2006introduction}. To ascertain the level of security an encryption algorithm offers, cryptanalysts will often employ cryptanalysis techniques to understand the limitations and weaknesses of an algorithm \cite{swenson2008modern}. In simple terms, the process of cryptanalysis is an attempt at breaking bad encryption algorithms to make good ones \cite{swenson2008modern}. This makes cryptanalysis a crucial component in the development of a robust encryption algorithm. However, the process of cryptanalysis is inefficient, consuming time, memory, and data \cite{easttom2021cryptanalysis}. As the size of the encryption algorithm increases, which is the data component, the amount of memory and time required will also increase. While existing state-of-the-art research employs heuristic methods such as Monte Carlo Search (MCS) to improve Differential Cryptanalysis (DC) efficiency \cite{dwivedi2019differential, dwivedi2023security}, heuristic methods depend on inefficient random selection that introduces a degree of imprecision to the cryptanalysis algorithm \cite{li1998heuristic, marnay1991effectiveness}. Further, as highlighted by the authors of \cite{dwivedi2023security}, despite the heuristic nature of their method producing state-of-the-art results, experimental results vary due to the random nature of the heuristic. This makes it difficult to reproduce results, which is counter to accepted scientific methods \cite{gigerenzer2011heuristic}. It is therefore necessary to reduce the negative influences of MCS heuristics on the cryptanalysis of lightweight encryption algorithms such as SIMON.

While DC has been identified as a powerful tool to analyse the security of LEAs such as SIMON, the size of SIMON makes the analysis difficult and inefficient. To address the large state space of SIMON, researchers have devised solutions utilising difference distribution tables (DDT). A DDT comprises a matrix that assists cryptanalysts or cryptographers in analysing how differences to an input propagate to differences in the output. The potential size of a DDT matrix, which depends on the number of inputs, represents a DDT limitation. The DDT for a n-bit S-box, which normally operates on a 4-bit or 8-bit word \cite{biryukov2014automatic}, will comprise a matrix of $2^{n} \times 2^{n}$. Constructing an S-box DDT would be relatively simple for small state sizes, however, addition, rotate XOR (ARX) cyphers such as SIMON, leverage modular addition to introduce non-linearity \cite{biryukov2014automatic}. Therefore, constructing a DDT for n-bit words is infeasible as it will require $2^{3n} \times 4$ bytes of memory for a standard 32-bit word size \cite{biryukov2014automatic}. To account for a large DDT of an ARX cypher such as SIMON, the authors of \cite{biryukov2014automatic} proposed using a Partial DDT (PDDT). The use of a PDDT, which is discussed in detail in Section \ref{SubSec:PDDT}, reduces the search space of the table while maintaining its effectiveness.

\begin{figure*}[t]
    \centering
    \includegraphics[width=0.65\linewidth]{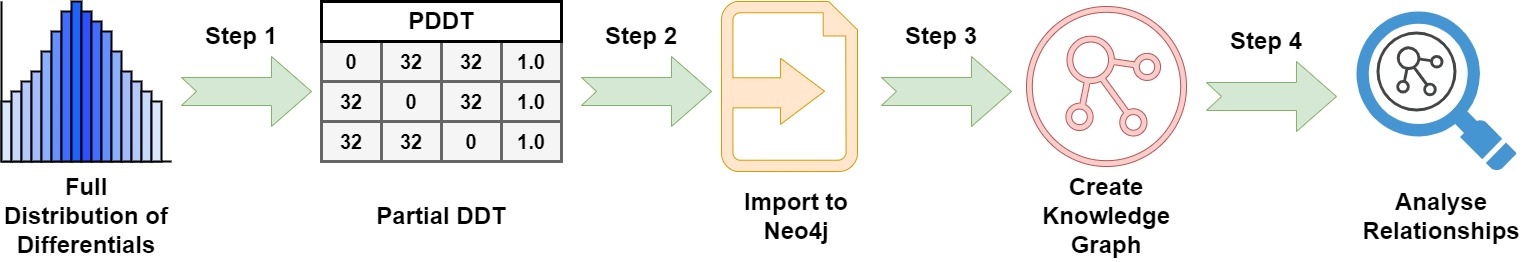}
    \caption{Our methodology: 1) \textit{Create a PDDT} from the full distribution of differentials. 2) \textit{Import}  the PDDT to Neo4j database. 3) \textit{Create knowledge graph} using Neo4j. 4) \textit{Analyse relationships} between differentials using Neo4j. }
    \label{fig:know_graph_methodology}
\end{figure*}

Like many data sources, PDDTs contain vast quantities of data. While the data in a PDDT may not appear statistically useful in an analytical sense, it is data, nonetheless. This aspect presents opportunities to explore the structure of the data. One such tool that is particularly powerful in identifying intricate relationships and patterns within data is a knowledge graph \cite{fensel2020introduction}. A knowledge graph integrates diverse data sources to create an interconnected network of entities that identifies relationships within the data \cite{ehrlinger2016towards}. This enables the exploration of relationships between entities, offering insights into data architecture and dependencies. Initially introduced by Google to enhance its search engine capabilities, knowledge graphs play a crucial role in many contemporary digital technologies \cite{fensel2020introduction, peng2023knowledge, wang2019explainable}. A key strength of knowledge graphs is their effectiveness at information retrieval, making them an optimal tool for inferring information about a cypher. Although fully decrypting an encrypted message using cryptanalysis methods is often infeasible, information deduction about the cypher is considered an effective goal of cryptanalysis \cite{easttom2021cryptanalysis}.

\subsection{Problem} \label{Subsec:Problem}
While PDDTs can be utilised to improve the efficiency of DC, current techniques are limited by heuristics. As highlighted above, state-of-the-art techniques utilising heuristic methods for cryptanalysis introduce a degree of inefficiency and imprecision which can inhibit attempts to reproduce experimental results. Despite the state-of-the-art MCS techniques producing relatively efficient results, the chosen path is not always the most efficient. A contributing factor is the inability of the MCS to efficiently calculate the most direct path as it relies on random path selection. While MCS can produce efficient results through heuristic methods, its inability to identify patterns and relationships inhibits its practicality to produce efficient, reproducible cryptanalysis. Knowledge graphs, however, are often used to identify patterns and intricate relationships within data, highlighting a shortcoming of MCS. While the power of knowledge graphs has been utilised in several domains, their application in cryptanalysis remains mostly untested, with only a preliminary analysis comparing node clustering between samples in the work by the authors of \cite{cook2024lightweight}. The shortcomings of state-of-the-art heuristic methods and the potential benefits of knowledge graphs in cryptanalysis raise the following thought-provoking research question. \textit{How can knowledge graphs be utilised to identify patterns and relationships between differentials in a PDDT?}

\subsection{Contribution} \label{Subsec:Contribution}
Having identified the limitations of existing state-of-the-art techniques, this paper presents a novel method for the efficient identification of paths through SIMON differentials utilising a Neo4j knowledge graph as illustrated in Figure \ref{fig:know_graph_methodology}. This paper forms the foundations for future research in applying knowledge graph-based machine learning (ML) to the DC of lightweight clock cyphers such as SIMON. 
To the best of our knowledge, this research presents a novel investigation of the relationship between cryptographic differential utilising Neo4j.

\subsection{Structure of the Paper}
The subsequent sections of this manuscript are organised in the following manner. Section \ref{Sec:Preliminaries} presents the preliminary background information necessary to understand the domain. It describes the SIMON LEA
and developing a partial difference distribution table. Section \ref{Sec:Methodology} describes the methods employed in this research including software selection and a description of the algorithms deployed. Section \ref{Sec:Results} presents the results and findings of this study. Finally, the conclusion and future works are presented in Section \ref{Sec:Conclusion}.

\section{Preliminaries} \label{Sec:Preliminaries}
This section describes the lightweight encryption algorithm, SIMON, calculating differential probabilities and PDDT. This section will provide the reader with knowledge of the domain necessary to understand the context of our contribution. Readers are referred to Table \ref{Tab:Symbols} for a description of the symbols and notations used in this study.

\begin{table}
\caption{Symbols and notations}
\centering
\scriptsize
\begin{tabular}{ll}
\hline
\textbf{Notation}      & \textbf{Description}  \\ \hline
$n$         & Number of bits \\ \hline
$m$         & The keyword size of either 2, 3, 4  \\ \hline
$\land$          & Bitwise AND \\ \hline
$\lor$          & Bitwise OR    \\ \hline
$\oplus$        & Bitwise exclusive OR (XOR) \\ \hline
$\boxplus$        & Modulo addition \\ \hline
$ \xrightarrow{n} $        & Right shift by $n$ bits \\ \hline
$ \xleftarrow{n}$        & Left shift by $n$ bits \\ \hline
$k_{(i)}$        & Key $i$ round \\ \hline
$\leq$        & Less than or equal to \\ \hline
$p_{\beta}(b)$ & Probability density function of $\beta$   \\ \hline
$w$         & Hamming weight \\ \hline
$h*(x)$         & Count of non-zero bits in $x$ \\ \hline
$DP$             & Differential probability \\ \hline 
$P_{th}$             & Pre-defined threshold \\ \hline 
$\neg$          & Logical negation operator \\ \hline
\end{tabular}

\label{Tab:Symbols}
\end{table}

\subsection{Description of the SIMON Cypher}\label{SubSec:SimonCypher}
In 2013, the NSA unveiled SIMON and SPECK, two innovative Feistel block cyphers, tailored specifically for environments with limited computational resources \cite{beaulieu2015simon}. SIMON was designed for hardware-constrained devices, such as IoT. The SIMON LEA incorporates five variants of $2n$-bit states, where $n$ represents the word size. With $n = 16, 24, 32, 48, 64$, supporting a block size of $32, 48, 64, 96$ and $128$ bits. SIMON comprises key sizes of $m \times n$ bit words. $m$ can be $2, 3, $ or $4$ and is based on the size of $n$. The following rules define the size of $m$ \cite{beaulieu2013simon}. \textit{If $n$ equals $16$ then $m$ must be $4$. If $n$ equals $24$ or $32$ then $m$ may be $3$ or $4$. $m$ may be $2$ or $3$ if $n$ equals $48$. $m$ can be $2, 3$ or $4$ if $n$ equals $64$.}

\begin{figure}
    \centering
    \includegraphics[width=0.55\columnwidth]{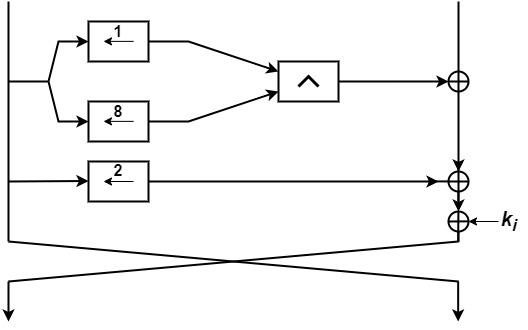}
    \caption{SIMON round function}
    \label{fig:simon_round_function}
\end{figure}

We can represent SIMON as follows with the parameters presented. SIMON $2 n / mn$ has a block size of $2n$ bits and a key size of $m \times n$ bits \cite{beaulieu2013simon}. As an example, SIMON32/64 is the version of SIMON with 32-bit plaintext blocks using a 64-bit key of $4(m) \times 16(n)$, where $n$ is the word size. SIMON, as with all Feistel block cyphers, utilises round functions for the encryption and decryption process. The SIMON round function uses a bitwise XOR $(\oplus)$, bitwise AND ($\land$) and left circular shift $(\xleftarrow{i})$ by $i$ bits \cite{beaulieu2013simon}. The SIMON round function shown in Figure \ref{fig:simon_round_function} can be defined as:

\begin{equation}
    f(x) = \left( \xleftarrow{1} x \right) \land \left( \xleftarrow{8} x \right) \oplus \left( \xleftarrow{2} x \right) .
\end{equation}

\subsection{Calculating Differential Probabilities}\label{SubSec:CalcDiffProb}
Differential probability refers to the probability of a random variable falling within a range of a specified value. According to the authors of \cite{dwivedi2019differential}, the input and output differences with the differences internally following each round define the differential characteristics. As such, the following equations define conditions for valid differentials and a method to calculate their hamming weights \cite{dwivedi2019differential}. The calculations are crucial in assessing the strength of a cryptographic cypher against DC.

Define $xdp^\land(a, b \rightarrow c)$ as the probability associated with the XOR-based differential operations in the context of addition modulo $2n$, where $a$ and $b$ are the input differences and $c$ is the output difference. The authors in \cite{dwivedi2019differential} state, differential $(a, b \rightarrow c)$ holds true exclusively under the condition if: 

\begin{equation} 
    eq(a \xleftarrow{1}, b \xleftarrow{1}, c \xleftarrow{1}) \!\wedge \! (a \oplus b \oplus c \oplus (b \xleftarrow{1}))\!=\!0
\end{equation}

where

\begin{equation} 
    eq(p,q,r):= (\neg p\oplus q) \wedge (\neg p \oplus r)
\end{equation}

The hamming weight ($w$), defined as $w(a, b \rightarrow c)$, for every valid differential of $(a, b \rightarrow c)$ can be defined as:

\begin{equation} 
    w(a, b \rightarrow c)= -\log _{2}(xdp^{+}(a, b \rightarrow c))
\end{equation}

The calculation of the hamming weight ($w$) for a legitimate differential employs the following equation:

\begin{equation} 
    w(a, b \rightarrow c):= h^{*}(\neg eq(a, b \rightarrow c)),
\end{equation}

where $h^{*}(x)$ is the sum of bits in $x$ that are not zero, excluding $x[n - 1]$.

\subsection{Partial Difference Distribution Tables}\label{SubSec:PDDT}
 
As highlighted above, a DDT for an LEA such as SIMON is infeasibly large, so to address this problem the authors of \cite{biryukov2014automatic} proposed using a PDDT. A PDDT represents a table where all XOR differentials of $(a, b \rightarrow c)$ are recorded on the condition that the differential probabilities (DP) equals or exceeds a pre-determined threshold ($P_{th}$) and can be expressed as:  

\begin{equation} 
    (a,b,c) \in PDDT \Leftrightarrow DP(a, b \rightarrow c) \geq P_{th}
\end{equation}

The efficient computation of the PDDT can be performed with the following proposition: 

\begin{prop}
The DP of the XOR operation combined with addition modulo $2n$ diminishes as the word size of the differences $a$,$b$,$c$ increases, such that:

\begin{equation} 
    p_{n} \leq....... \leq p_{k} \leq p_{k-1} \leq.... \leq p_{1} \leq p_{0},
\end{equation}
\end{prop}

where $pk$ represents the DP from $a_k, b_k$ progressing to $c_k$, where $n \geq k \geq 1,p_0=1$ and $x_k$ signifies the $k$ least-significant bit (LSB) difference $x$, such that $x_k=x[k-1:0]$. The algorithm constructs the PDDT beginning at the LSB position $k=0$. It then assigns the values to $a_k$,$b_k$ and $c_k$ recursively \cite{biryukov2014automatic}. At each bit position $k$, for $n > k \geq 0$, the likelihood of a $k + 1^{th}$ bit differential with a partial structure is evaluated to ascertain whether it exceeds the specified threshold. If the evaluation returns true, the algorithm proceeds to the next bit. If it evaluates as false, the algorithm will return and assign different values to $a_k,b_k$ and $c_k$. The algorithm repeats the procedure and is terminated when $k=n$. At this point, the XOR differential $(a_k,b_k \rightarrow c_k)$ will be added to the PDDT. $k$ is initialised with a value of $0$ and $a_0 = b_0 = c_0 = \Phi$ \cite{biryukov2014automatic, dwivedi2019differential}.

As identified by both \cite{biryukov2014automatic} and \cite{dwivedi2019differential}, creating a PDDT is time-consuming and resource-intensive. To expedite the process of creating a PDDT and increase time efficiency, the authors recommend adjusting the tolerance of the threshold value. The lower the value, the larger the PDDT and the longer the table takes to create. Both authors record a threshold probability of 0.1 creating 3951388 elements in 1 minute and 29 seconds. If the threshold probability is decreased to 0.06 the search space increases, producing a table with 167065948 elements in 44 minutes and 36 seconds. The table size and time required grow exponentially as the threshold nears zero.

\section{Methodology} \label{Sec:Methodology}
This section presents the methods utilised to create a knowledge graph to identify relationships between differentials in the SIMON cypher. Our method is summarised in Figure \ref{fig:know_graph_methodology} highlighting the key steps in the process.

\subsection{PDDT and Comma Separated Values File Generation} \label{SubSec:CreatePDDT}
As highlighted in Section \ref{SubSec:PDDT}, developing a DDT is infeasible for the SIMON cypher due to its large search space. To counter this limitation and track changes within the differentials, a PDDT will be employed. Increasing the threshold probability reduces the search space, creating a more manageable PDDT that can still produce traceable changes to the differentials. This research will utilise Visual Studio IDE \cite{VisualSt22online} to create the PDDT. Visual Studio IDE will facilitate the creation of four text files generated by the C++ code provided by \cite{biryukov2014automatic}. While the PDDT generates four text files, Neo4j requires using a comma-separated values (CSV) file to import the data. The four text files will be imported into Python and transformed into a CSV file to accommodate this problem. As the generated CSV file is extremely large with many hundreds of thousands of rows, a sample of the differentials will be uploaded for analysis in Neo4j. The sample will be a quota-based sample of differentials containing 3 \% and at least one of each type of output differential. Due to the limited available space, we will skip the description of the process and algorithm associated with importing the CSV data to Neo4j.

\subsection{Knowledge Graphs} \label{SubSec:Neo4jGraph}
Creating a knowledge graph to study the complex relationships between differentials in cryptography requires using an appropriate tool to conduct the study, such as Neo4j \cite{Neo4j2020}. A particular benefit of Neo4j is its efficient streamlined approach to identifying connections between data points \cite{Neo4j2021What}. While programming languages such as Python with NetworkX \cite{hagberg2008exploring} process the connections at the time of query, Neo4j stores the connections directly in the database, significantly increasing performance. This study will use Neo4j's free version, allowing unrestricted replication of our results.

In the state-of-the-art work by the authors of \cite{dwivedi2023security}, the MCS goal is to reach a target hamming weight as efficiently as possible with an initial weight set very high. The weight is reduced when an explored branch weight exceeds the previous. Differentials with a high weight will therefore impose a significantly higher reduction on the weight than differentials with a weight nearer to zero. A cursory examination of the weights generated by the PDDT illustrates that weights range from $0.125$ to $1$ with a low proportion of differentials with a weight of $1$. As the weight approaches zero the representation of that weight increases within the list of differentials. This representation of the distribution of the weight data highlights the difficulty associated with heuristic exploration. To extract the most efficient path through the differentials, the traversal of short paths containing differentials with high weight is a priority as it will reduce the total number of steps necessary to converge to the ideal solution. 

As highlighted in Section \ref{Sec:Introduction}, a key benefit of graph databases is the ability to identify intricate relationships between data in a dataset. To facilitate identifying relationships between differentials, a Cypher query, as illustrated in Algorithm \ref{Alg:weightOutputRel}, is executed. This query will identify relationships between differentials based on the weight and output differential. As the existing state-of-the-art by the authors of \cite{dwivedi2023security} produces superior results from higher valued weights, limiting the selection of nodes to a weight greater than $0.5$ will create relationships that yield higher weights. However, due to complex relationships between differentials, weights below $0.5$ will be present in the final graph. Algorithm \ref{Alg:weightOutputRel} took $1240$ms to execute. Although this query only forms the relationships between differentials, it demonstrates significant efficiency gains compared to creating the PDDT as highlighted in Section \ref{SubSec:CreatePDDT}. Further, when compared to the DC of SIMON 32 by the authors in \cite{dwivedi2023security} and \cite{cook2024lightweight}, the time taken to identify the relationships is up to $96$ \% faster than the DC of the SIMON cypher. While the process of DC is still to be undertaken, identifying relationships and efficient paths of traversal can expedite the DC process.

\begin{algorithm}
	\caption{Create weight and output relationships in Neo4j} 
    \label{Alg:weightOutputRel}
	\begin{algorithmic}[1]
    \STATE MATCH: (a:DIFFERENTIALS),(b:DIFFERENTIALS)
    \STATE WHERE:  a.output $\leq 0$ AND b.weight $\leq 0.5$
    \STATE CREATE: $(a)-[r:OUTPUT\_WEIGHT]->(b)$
    \STATE RETURN: r
	\end{algorithmic} 
\end{algorithm}

Once the relationships have been formed, the next step is visualising the graph, as shown in Algorithm \ref{Alg:fullDiffRel}. This will allow for the deep analysis and study of relationships between differentials. The number of relationships displayed in Algorithm \ref{Alg:fullDiffRel} may be modified as required. It took Neo4j $124$ms to execute the command.

\begin{algorithm}
	\caption{Visualise full graph of differential relationships} 
    \label{Alg:fullDiffRel}
	\begin{algorithmic}[1]
        \STATE MATCH: p=()-[:OUTPUT\_WEIGHT]-() 
        \STATE RETURN: p 
        \STATE LIMIT: 2000
	\end{algorithmic} 
\end{algorithm}

\section{Results} \label{Sec:Results}
This section describes the results of the experiments to create a knowledge graph of the SIMON cypher PDDT. This study will focus on identifying relationships between differentials, referred to as nodes in this analysis. It will investigate the complexity and distance between nodes, and the optimal path traversal between nodes. The experiments in this section are deployed on a free Neo4j AuraDB instance on the Google Cloud Platform.

\begin{figure}
    \centering
    \includegraphics[width=0.55\linewidth]{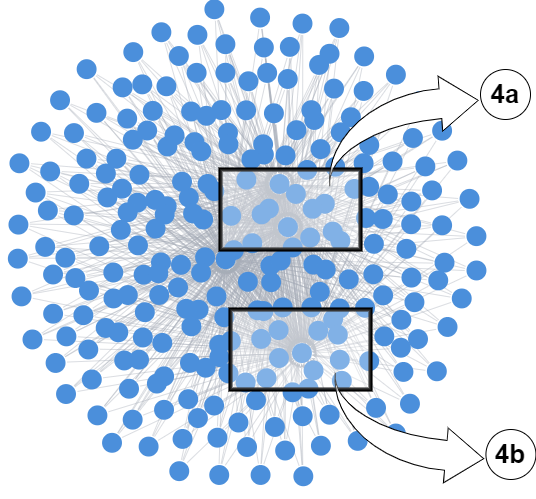}
    \caption{240 nodes connected by 960 relationships in Neo4j. Enlarged views of \ref{fig:zoom_a} and \ref{fig:zoom_b} are presented below. }
    \label{fig:full_graph}
\end{figure}

\begin{figure}[t]
  \centering
  \begin{subfigure}{.4\columnwidth}
    \centering
    \includegraphics[width=\linewidth]{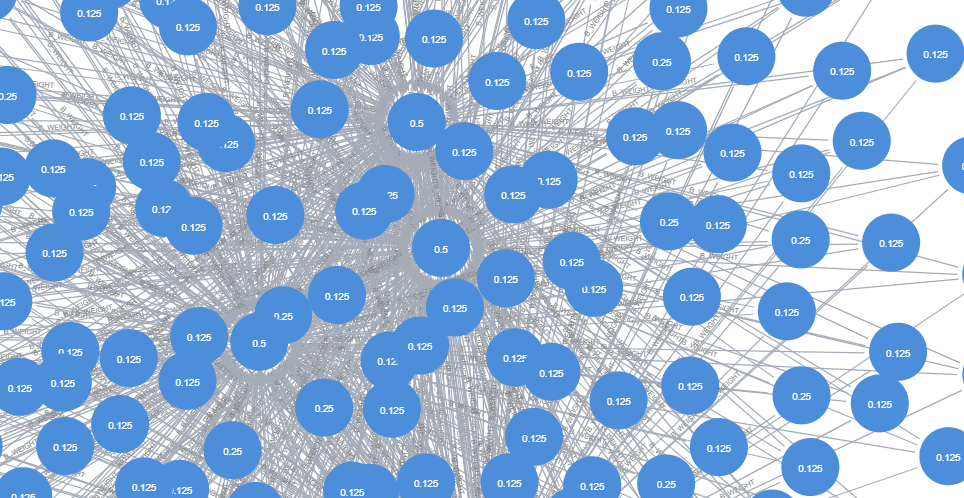}
    \caption{Enlarged view of neighbourhood \ref{fig:zoom_a} from Figure \ref{fig:full_graph}.}
    \label{fig:zoom_a}
  \end{subfigure}%
  \hfill
  \begin{subfigure}{.4\columnwidth}
    \centering
    \includegraphics[width=\linewidth]{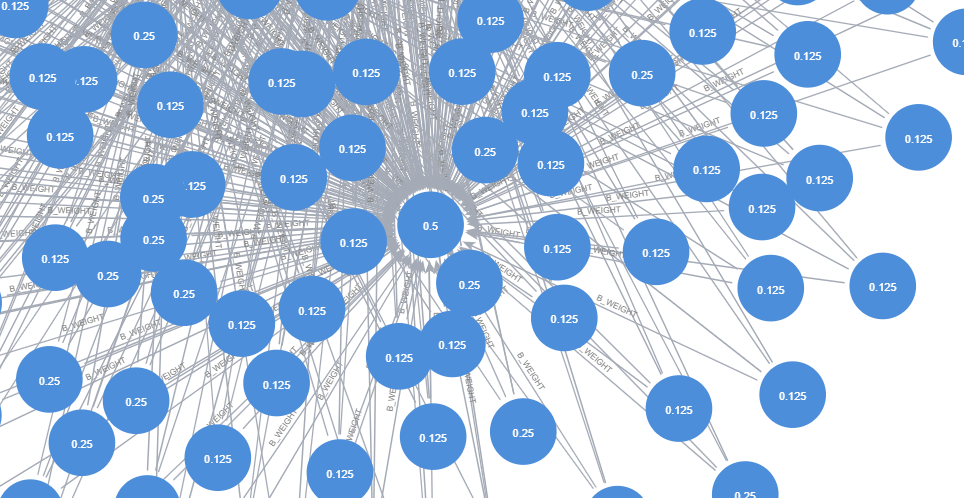}
    \caption{Enlarged view of neighbourhood \ref{fig:zoom_b} from Figure \ref{fig:full_graph}.}
    \label{fig:zoom_b}
  \end{subfigure}%

  \caption{Enlarged views of neighbourhoods \ref{fig:zoom_a} and \ref{fig:zoom_b}. }
\end{figure}

An observation of the full distribution of nodes and their relationships illustrated in Figure \ref{fig:full_graph} shows intense clustering around the $4$ centrally located nodes with a weight of $0.5$. As illustrated, most nodes are also located centrally, with nodes sharing multiple relationships. As the nodes extend towards the periphery, the lengths of the paths become greater and the weights of the nodes reduce, demonstrating highly inefficient paths. Although several nodes in the outer regions may connect to central nodes and be necessary paths of traversal, the goal is to avoid such paths if possible and limit their use.  While the full distribution illustrates a larger view of node relationships, this analysis will focus on differences and opportunities between regions \ref{fig:zoom_a} and \ref{fig:zoom_b}.

An investigation of Figure \ref{fig:zoom_a} illustrates a close clustering of nodes with a high weight and short path between nodes. As highlighted above, nodes with a high weight exhibit a more significant impact on the weight than a node with a smaller weight. Figure \ref{fig:zoom_a} contains $3$ of the $4$ nodes in the full distribution with a weight of $0.5$, and it similarly contains a high proportion of nodes with a weight of $0.25$, many with short paths. However, despite these promising aspects, several nodes of high weight have a long traversal from their closest highly weighted node, demonstrating a potentially inefficient path. Further, with a high proportion of nodes with a weight of $0.125$, traversal through these nodes, despite being necessary, could present challenges in optimal path selection.

A study of Figure \ref{fig:zoom_b} highlights several factors that inhibit the effectiveness of path traversal in this region. As illustrated, this region contains far fewer nodes and relationships than shown in Figure \ref{fig:zoom_a}, with many paths exhibiting a greater length than those in Figure \ref{fig:zoom_a}. Despite the proximity of three nodes with a weight of $0.25$ in this region, many other nodes possess either a low weight or are a great distance from the core node that weights $0.5$. This illustrates an inefficient area of path exploration, despite the area demonstrating a high cluster of nodes in FIgure \ref{fig:full_graph}.

\begin{figure}[t]
  \centering
  \begin{subfigure}{.275\columnwidth}
    \centering
    \includegraphics[width=\linewidth]{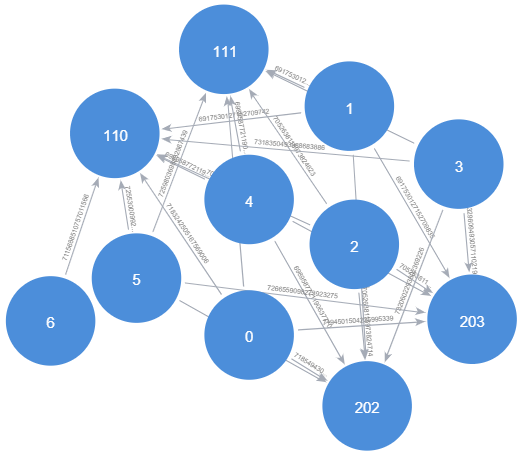}
    \caption{Nodes represented by unique ID.}
    \label{fig:nodes_by_id}
  \end{subfigure}%
  \hfill
  \begin{subfigure}{.275\columnwidth}
    \centering
    \includegraphics[width=\linewidth]{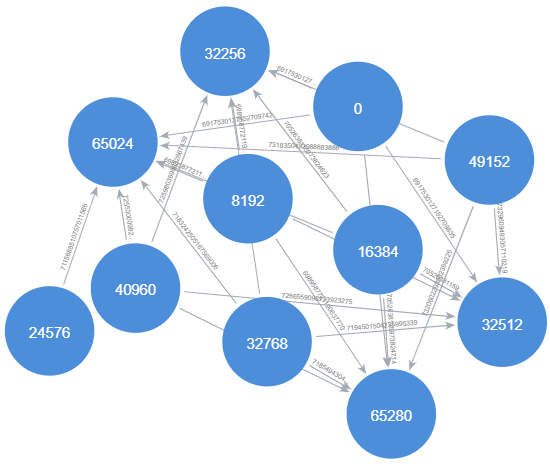}
    \caption{Nodes represented by output differential.}
    \label{fig:nodes_by_output}
  \end{subfigure}%
  \hfill
  \begin{subfigure}{.275\columnwidth}
    \centering
    \includegraphics[width=\linewidth]{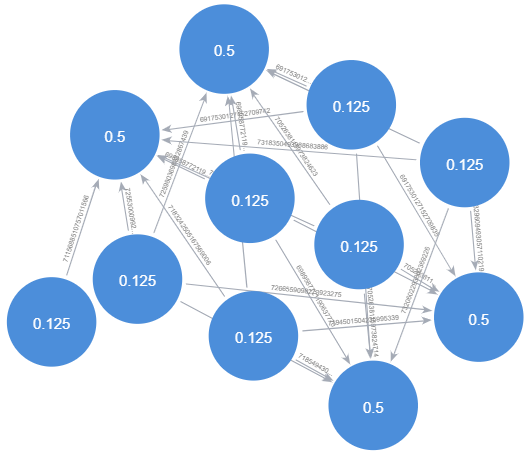}
    \caption{Nodes represented by weight.}
    \label{fig:nodes_by_weight}
  \end{subfigure}%
  \caption{The same knowledge graph com ID in Figure \ref{fig:nodes_by_id}, output differential in Figure \ref{fig:nodes_by_output} and weight in Figure \ref{fig:nodes_by_weight}. }
\end{figure}

While the investigation of two regions of a wider graph illustrates potential opportunities for exploration, it is necessary to identify specific nodes and paths for traversal. As an examination of the CSV file illustrates, several output differentials contain the same value with different associated inputs and weights. A reliance on the output value alone will result in an incorrect node selection. It is therefore necessary to identify the correct node and path for traversal. To accomplish this, Neo4j provides a unique identification (ID) number for each node and relationship pair within the graph, as illustrated in Figure \ref{fig:nodes_by_id}. By utilising the available tools in the Neo4j browser, it is possible to identify the unique node ID, its output differential, as shown in Figure \ref{fig:nodes_by_output}, weight as illustrated in Figure \ref{fig:nodes_by_weight} and the optimal paths to related nodes. While Neo4j permits using these labels for nodes and relationships, it is impossible to display the ID, weight, and output as a caption in the visualisation. However, the caption may be changed by selecting the property and caption to display. This action does not alter the relationships between nodes but allows for identifying unique nodes and paths within the graph.

\begin{figure}
    \centering
    \includegraphics[width=0.6\linewidth]{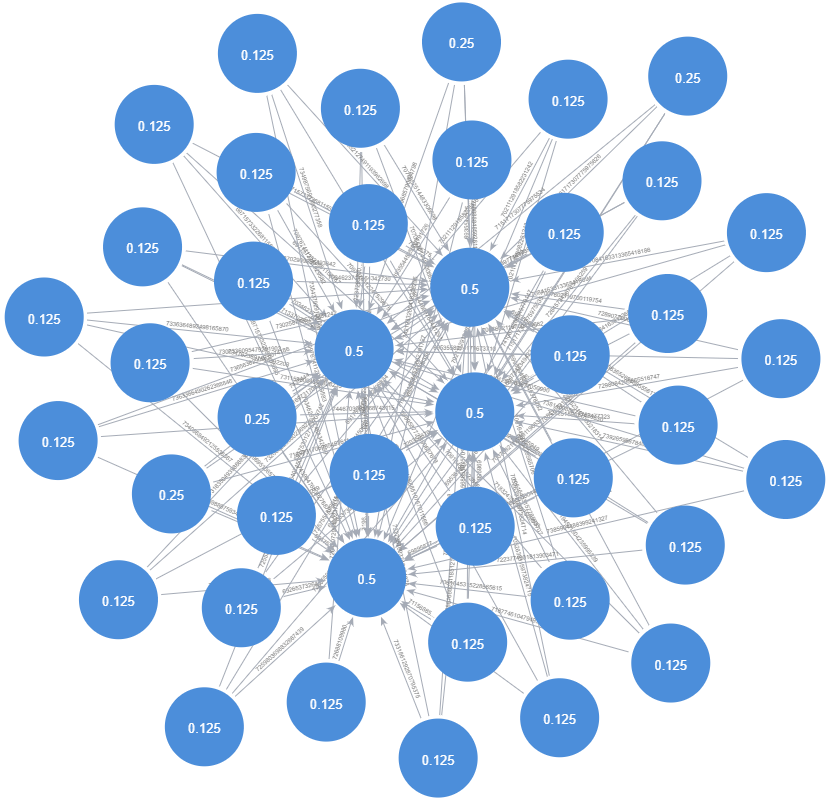}
    \caption{Neo4j Graph showing 150 relationships between nodes}
    \label{fig:neo4j_150_rels}
\end{figure}

A crucial factor to assess in the graph is the distance between nodes. Inspecting a large and complex graph, such as that in Figure \ref{fig:full_graph}, is difficult due to the large number of nodes and connecting relationships. Reducing the number of relationships in the graph makes it less complicated to inspect the relationships and identify the path distances, as illustrated in Figure \ref{fig:neo4j_150_rels}. While cryptanalysis of a small selection of nodes is infeasible, it highlights the complex nature of differential relationships within a PDDT. As observed, an optimal path between nodes with high weights may traverse multiple nodes with low weights. Similarly, some paths may be shorter than those with similar total weights. The optimal path will be the shortest distance, highest weight and least number of nodes.

\begin{figure}
    \centering
    \includegraphics[width=0.55\linewidth]{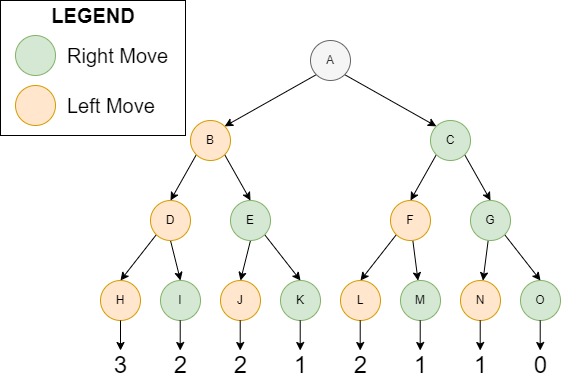}
    \caption{MCS method utilised by \cite{dwivedi2023security} and \cite{cook2024lightweight}}
    \label{fig:nmcs_search}
\end{figure}

The existing state-of-the-art methods by the authors of \cite{dwivedi2023security} and \cite{cook2024lightweight} utilise MCS to determine the best path throughout the differentials. While the heuristic can produce efficient results, the random nature can also be inefficient. As illustrated in Figure \ref{fig:nmcs_search}, the MCS method utilises a tree-like structure for the random investigation of paths to find the path with the lowest weight, however, the MCS does not know the structure of the tree. As shown, there is a 50 \% chance of navigating a path with a weight of 1 or less. The heuristic may traverse several undesirable paths before identifying the optimal path, impeding search efficiency. Conversely, our method identifies all optimal paths and differentials, allowing for efficient navigation through the differentials. Further, the existing methods investigate the path selection at the time of execution, which takes more than 10 seconds to execute. Conversely, our proposed Neo4j method identifies relationships in 1240ms and produces the visualised graph in 124ms. This demonstrates a reduction of up to $96$ \% in the time required to identify paths.

\section{Conclusion and future research }\label{Sec:Conclusion}

In this paper, we have demonstrated the innovative application of Neo4j graph database to analyse the PDDT of the SIMON cypher, which offers new insights into the complex relationships and connections within the cryptographic structure. By leveraging the powerful capabilities of Neo4j, we could effectively model and explore intricate networks of nodes representing the PDDT, which allowed for a deeper understanding of the cyphers characteristics. Our analysis examined the relationships between nodes, comparing regions of node clusters, with an additional investigation of the hamming weight and distance between them. This approach allows for the detection and exploration of patterns and relationships that were not readily evident, highlighting the utility of Neo4j graph databases in cryptanalysis. The comparison of regions revealed distinct characteristics in the distribution of nodes, weights, and the efficiency of connecting paths. The application of Neo4j enriches the understanding of the SIMON cypher and demonstrates the broader potential of graph databases in cryptanalysis. Further, the simple and efficient use of Neo4j can be applied to other cyphers, expanding our knowledge and capabilities in securing critical information. Future studies may investigate the application of ML to use the information gathered through a knowledge graph, allowing for the identification of critical vulnerabilities within cyphers.

 \nocite{*} 
\bibliographystyle{unsrt}
\bibliography{references}  






\end{document}